**Optical properties of spin coated Cu doped ZnO nanocomposite films**

P. Samarasekara and Udumbara Wijesinghe

Department of Physics, University of Peradeniya, Peradeniya, Sri Lanka

**Abstract**

Spin coating technique was used to synthesize pure ZnO and Cu doped ZnO films on amorphous and conducting glass substrates. The doped amount of Cu in ZnO was varied up to 5% in atomic percentage. Speed of spin coating system, coating time, initial chemical solution and annealing conditions were varied to optimize the properties of samples. Transmittance of samples was measured for ZnO doped with 1, 2, 3, 4 and 5% of Cu. Absorbance, reflectance and refractive index were derived from the measured transmittance. Film thickness of each film was calculated using the graphs of refractive index versus wavelength. Film thickness varies in a random manner depending on the amount of ZnO or Cu doped ZnO solutions spread on the substrate. The energy gap of each film was calculated using the graph of square of absorption coefficient time photon energy versus photon energy. The calculated energy gap values of Cu doped ZnO film samples decrease with the Cu concentration in ZnO. This means that the conductivity of ZnO can be increased by adding a trace amount of conducting material such as Cu.

**1. Introduction:**

ZnO films are prime candidates of photocells, gas sensors, electronic devices, optoelectronic devices, acoustic wave devices and piezoelectric devices. ZnO indicates transparent properties due to its high band gap of 3.2 to 3.3 eV. As a result, ZnO is used to absorb UV part of the solar spectrum. Most of the oxides including ZnO are used as gas sensors. The resistivity of undoped ZnO varies from $10^{-6}$ to $10^{6}$ $\Omega$m depending on the preparation technique. However, the resistivity can be decreased by doping with a conducting materials or metals. ZnO films have been synthesized on glass substrates using spray pyrolysis method [1], sol-gel process [2] and spin coating method [3]. Also thin films of ZnO have been prepared using dc and rf sputtering [4] and on Si(111) substrates using pulsed laser deposition (PLD) [5]. In addition, highly aligned ZnO films have been deposited [6]. Stability of cobalt doped ZnO films with deposition temperature has been investigated [7]. Transparent conductive ZnO films doped with Co and In have been fabricated on glass substrates at 350 $^{o}$C using ultrasonic spray method [8]. According to X ray diffraction (XRD) patterns, these films have indicated a (002) preferential direction. After doping with Co, the band energy gap of these films has increased from 3.25 to 3.36 eV. After doping with In, the band energy gap of these films has decreased from 3.25 to 3.18 eV. In and Co have been added to ZnO to enhance the conductive properties of ZnO [8].

Variation of structural and electrical properties with thickness of Ga doped ZnO films prepared by reactive plasma deposition has been investigated [9]. Li, P and N doped ZnO thin films have been grown using PLD [10]. Fourier transform infrared spectrometer (FTIR) and XRD properties of Al doped ZnO films deposited on polycrystalline alumina substrates by ultrasonic spray pyrolysis have been investigated [11]. Optical properties and photoconductivity of ZnO thin films grown by pulsed filtered cathodic arc vacuum technique have been studied [12]. Structural and optical properties of ZnO thin films synthesized on (111) $CaF_2$ substrates by magnetron sputtering have been investigated [13]. Effect of substrate temperature on the crystalline properties Al doped ZnO films fabricated on glass substrates by RF magnetron sputtering method has been studied [14]. Low temperature annealing effect on the structural and optical properties of ZnO films deposited by PLD has been investigated [15].

Previously ZnO films have been deposited using reactive dc sputtering method by us [16]. Photo-voltaic and absorption properties of ZnO thin films deposited at different sputtering and annealing conditions were investigated [16]. In this report, the optical properties of spin coated ZnO films have been explained. The energy gap, optical absorption, and thickness have been calculated for films fabricated at different rotational speeds of spin coated system for different coating times and films annealed at different temperatures for different time periods. The energy gap, optical absorption, and thickness solely depend on the Cu concentrations doped in ZnO films. The spin coating method is a low cost technique compared to the deposition techniques required vacuum or expensive equipments.

Low cost P-$Cu_2O$/N-CuO junction was prepared using thermal evaporation method [17]. Furthermore, multiwalled carbon nanotubes synthesized using chemical vapor deposition method was employed to detect $H_2$ and methane gases [18]. Copper oxide was fabricated using reactive dc sputtering by us [19]. Energy gap of semiconductor particles doped with salts were determined by us [20]. In addition, ZnO indicates some magnetic properties. Theoretical studies of magnetic films have been carried out using Heisenberg Hamiltonian [21-26].

## 2. Experimental:

Chemicals with purity higher than 98% were used as starting materials. Copper acetate dehydrate (Cu(CH3COO)$_2$.2H$_2$O), anhydrous ethanol and monoethanolamine (MEA), potassium iodide (KI), tetrapropylammonium iodide (Pr$_4$NI) and iodine were used without grinding. Zinc acetate dihydrate (Zn(CH3COO)$_2$.2H$_2$O) was grinded before use. All the chemicals except I$_2$ and ethanol were vacuum dried at 60 $^o$C for 24 hours prior to use.

ZnO solutions were prepared using zinc acetate dihydrate (Zn(CH3COO)$_2$.2H$_2$O), anhydrous ethanol and monoethanolamine (MEA) as the solute, solvent and sol stabilizer, respectively. Zinc acetate was first grinded and dissolved in a mixture of ethanol and monoethanolamine at room temperature. The molar ratio of MEA to zinc acetate was kept at 1:1. Five doped solutions were prepared by adding copper acetate dihydrate (Cu(CH3COO)$_2$.2H$_2$O) to the mixture with an atomic percentage of Cu varying from 1%  to 5%. The resulting solutions were stirred by a magnetic stirring apparatus at 70 ◦C for an hour. Finally transparent ZnO solutions were formed. In the sol, the Zn concentration was 0.5 mol/L. The prepared sols were aged for 24 hours at room temperature. Then the thin films were prepared by a spin-coating method on glass substrates which had been pre-cleaned by detergent, and then cleaned in methanol and acetone for 10 min each by using ultrasonic cleaner and then cleaned with deionized water and dried. The films samples were grown at 2000rpm for 30s. After coating, the sample was first dried at 200 ◦C for 10min, and then was annealed at 500 ◦C in ambient atmosphere for an hour.

The optical measurements of the films were carried out using Shimadzu UV 1800 spectrophotometer in the wavelength range from 190 to 900 nm at room temperature. Samples synthesized on amorphous insulator glass substrates were used to measure the optical properties.

### 3. Results and Discussion:

Experimental measurements are usually made in terms of percentage transmittance (*T%*), which is defined as,

$$T\% = \frac{I}{I_0} \times 100\% \tag{1}$$

where I is the light intensity after it passes through the sample and $I_o$ is the initial light intensity.

All the samples given in this report were prepared at 2000 rpm in 30s and subsequently annealed at 500 ◦C in air for an hour. After optimizing synthesize conditions, above conditions were selected to prepare samples. Figure 1 shows the transmittance curves for undoped ZnO films (solid line) and ZnO doped with 3 (dashed line) and 5% (dotted line) of Cu. The atomic doping percentages of Cu are given here. Although all the measurements were performed for doping concentrations of 1, 2, 3, 4 and 5% of Cu, only the curves for 3 and 5% are shown in this manuscript. The absorption edge is observed around 380 nm. Below the absorption edge, the transmittance increases with Cu concentration in ZnO thin film sample. However, just above the transmittance edge, pure ZnO film and film sample with 3% of Cu indicates the highest and lowest transmittance, respectively. At longer wavelengths, film sample with 3% of Cu dominates. Because different materials are capable to absorb different wavelengths, the dominance of transmittance varies with the wavelength even for the same concentration of Cu.

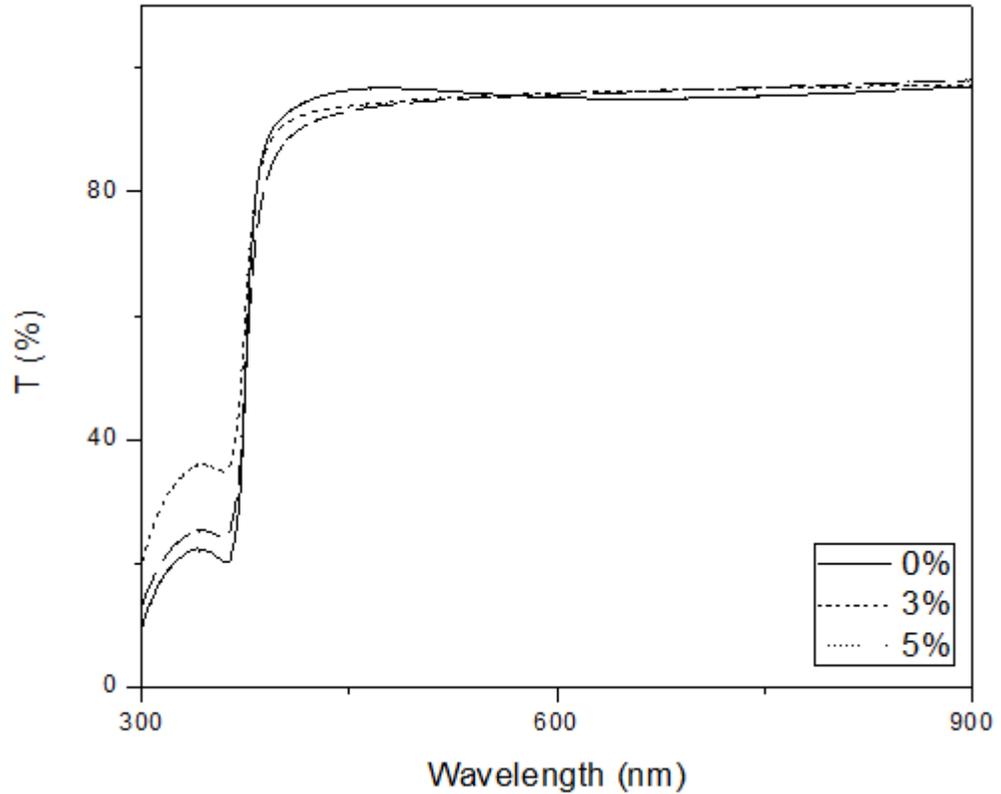

Figure 1: Transmittance versus wavelength for undoped ZnO and doping concentrations of 3 and 5% of Cu.

The relationship between absorbance ($A$) and transmittance ($T$) is given by,

$$A = -\log_{10}(T) = -\log_{10}\frac{I}{I_0} \qquad \textbf{(2)}$$

The curves of absorbance versus wavelength for undoped ZnO (solid line) films and ZnO doped with 3 (dashed line) and 5% (dotted line) of Cu are given in figure 2. The absorption edge can be observed again around 380 nm. Below this wavelength, absorption decreases with Cu concentration of ZnO sample. Just above this wavelength, sample with 3% of Cu indicates the highest absorption. This variation is obvious, because the absorption is the opposite phenomena of transmittance. Again the relative variation of absorption between samples with different concentrations is due to the reason explained previously.

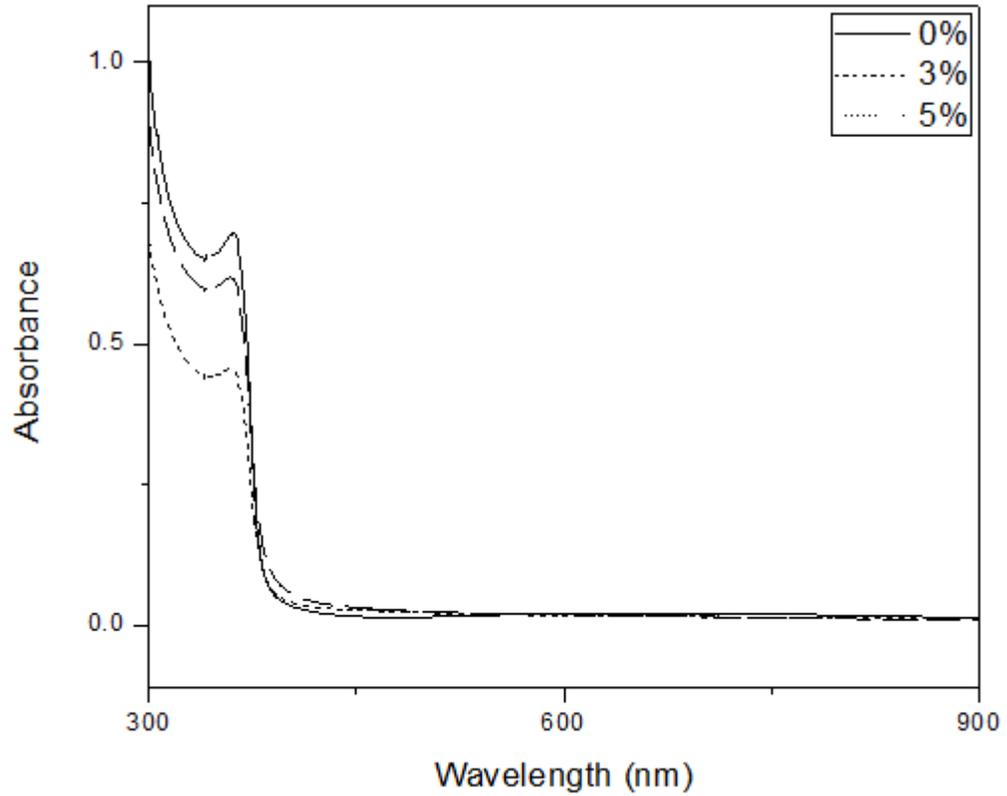

Figure 2: Absorbance versus wavelength for undoped ZnO sample and Cu doping concentrations of 3 and 5%.

The reflectance ($R$) was calculated using the following relation

$$R = 1 - (T \times e^A)^{0.5} \qquad\qquad (3)$$

where R is the reflectance, T is the transmittance and A is the absorbance.

Figure 3 shows the graph between reflectance and wavelength for pure ZnO and Cu concentrations of 3 and 5%. The curves observed in reflectance graph have some resemblances to the absorbance curves.

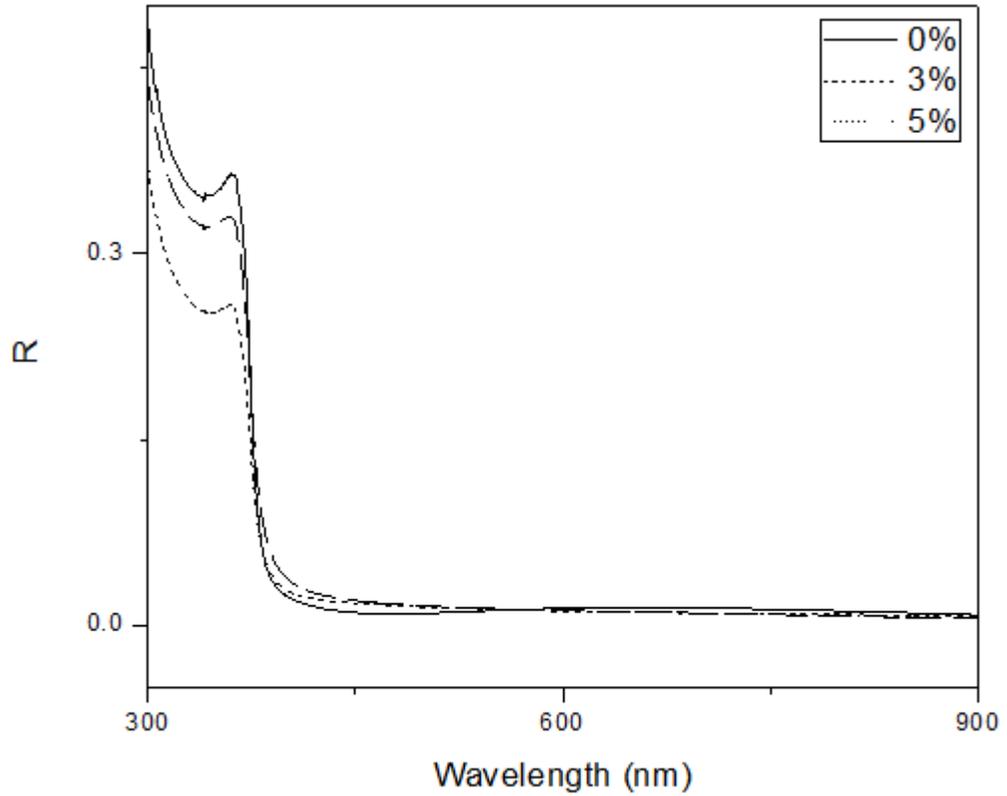

Figure 3: Reflectance versus wavelength for undoped ZnO and Cu doped samples.

The refractive index at different wavelengths were calculated using equation,

$$n = \left( \frac{1 + R^{0.5}}{1 - R^{0.5}} \right) \qquad (4)$$

The graph between refractive index and the wavelength is given in figure 4 for pure ZnO and Cu doped samples. Variation of refractive index is similar to the variation of absorbance and reflectance.

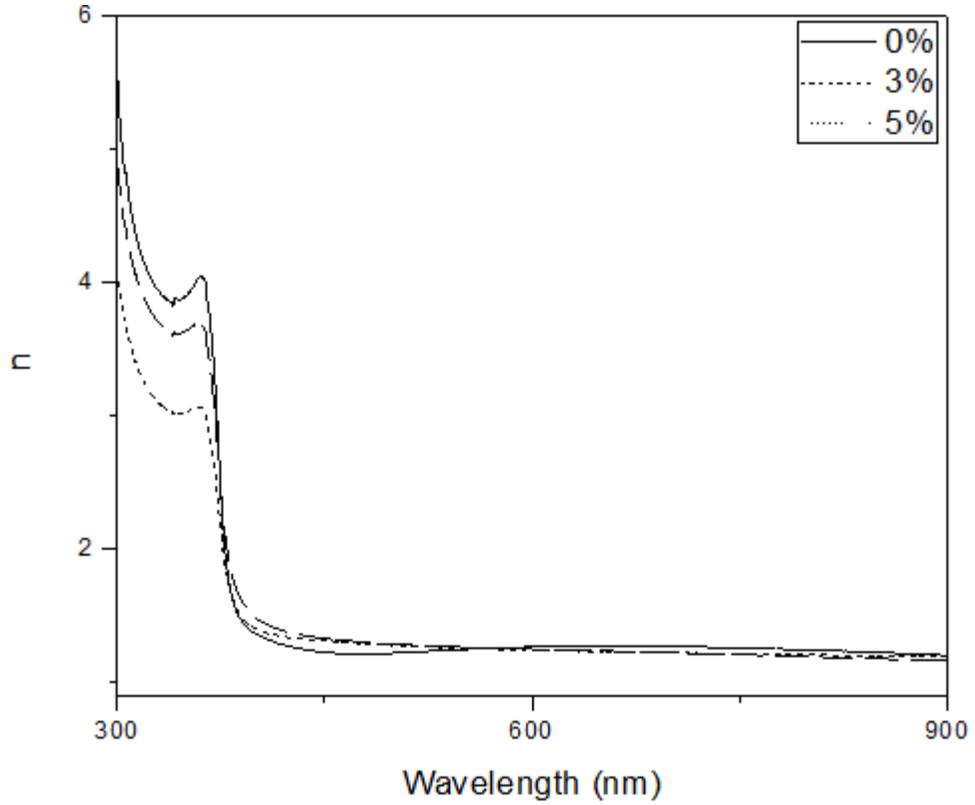

Figure 4: Refractive index versus wavelength for undoped ZnO and Cu doped samples.

The thickness of the thin film can be calculated by equation

$$t = \frac{\lambda_1 \lambda_2}{(n_1 \lambda_2 - n_2 \lambda_1)} \qquad (5)$$

where $\lambda_1$ and $\lambda_2$ are the wavelengths at which two successive maxima or minima occur, and $n_1$ and $n_2$ are the corresponding refractive indices.

The thicknesses calculated for the films with different Cu concentrations using equation (5) and figure 4 are given in table 1. Because the thickness of film depends on the initial amount of solution spread on the substrate, the thickness varies from film to film in a random manner as given in table 1.

| Cu | $\lambda_1$/nm | $\lambda_2$/nm | $n_1$ | $n_2$ | d/nm |
|---|---|---|---|---|---|
| 0% | 354.6 | 487.1 | 3.96448 | 1.2136 | 115.0925 |
| 1% | 359 | 899.9 | 3.76367 | 1.12466 | 108.2954 |
| 2% | 358.2 | 899.6 | 3.75074 | 1.1681 | 109.0202 |
| 3% | 360.6 | 899.4 | 3.50546 | 1.17383 | 118.8204 |
| 4% | 359.1 | 899.6 | 3.70123 | 1.16073 | 110.9055 |
| 5% | 358.2 | 899.6 | 3.07063 | 1.19296 | 138.0017 |

Table 1: Film thicknesses of Cu doped ZnO films.

The fundamental absorption, which corresponds to electron excitation from the valence band to conduction band, can be used to determine the value of the optical band gap. The absorption coefficient α was determined using the relation

$$\alpha = 2.303 \times \left(\frac{A}{t}\right) \tag{6}$$

α for each sample at different wavelengths was calculated using figure 2 and table 1. The relationship between the absorption coefficient (α) and the incident photon energy (hυ) can be written as

$$\alpha h\nu = B(h\nu - E_g)^{0.5} \tag{7}$$

Where, $B$ is a constant, $E_g$ is the band gap energy of the material. When $\alpha h\upsilon$ is zero, hυ =$E_g$ from equation (7). Figure 5 shows the graph of $(\alpha h\upsilon)^2$ versus hυ for films with ZnO (solid line) and Cu concentrations of 3 (dashed line) and 5% (dotted line). The value of optical band gap was calculated by extrapolating the straight line portion of $(\alpha h\upsilon)^2$ versus hυ graph to hυ axis. According to above equation, the value of x axis at intercept of this straight line is the energy gap ($E_g$).

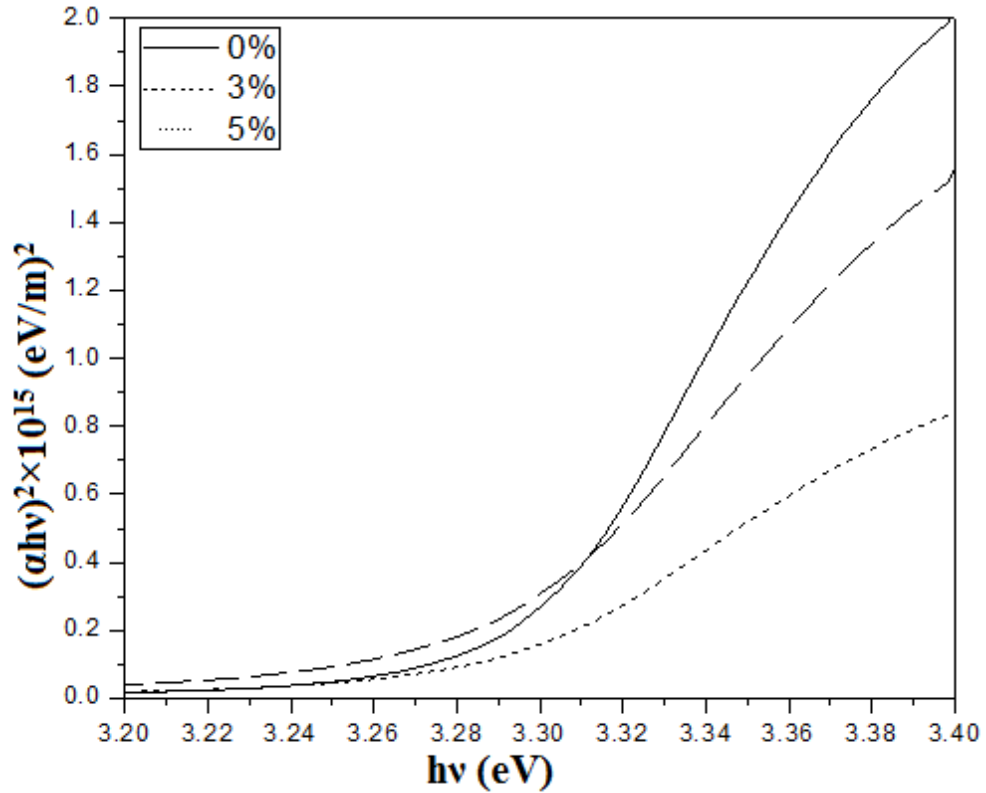

Figure 5: Graph of (αhν)² versus photon energy (hν).

Energy gaps calculated from figure 5 and equation (7) are given in table 2. Although only three curves are given in figure 5, the energy gaps calculated for all the concentrations are given in table 2. Band gap gradually decreases with the doped amount of Cu. The impurity energy levels and defects always contribute to the decrease of energy gap. This implies that the addition of Cu enhances the conductance of the sample as expected. However, this variation of energy gap (from 3.27 to 3.18 eV) is really small because a trace amount of Cu is added. By adding a trace amount of Cu, the conductivity of ZnO can be improved without altering the other properties of ZnO. Similar variation has been observed for indium doped ZnO films by some other researchers [8]. Electron concentration dependence of the band gap shift in Cu doped ZnO could be the reason for this decrease of energy gap. The increase of Urbach energy is the other possibility. The energy gap of our undoped ZnO film is really close to the standard band gap value of ZnO (3.2 to 3.3 eV). This confirms the formation of ZnO phase in thin film. Our synthesized films were apparently transparent. Diffraction peaks couldn't be observed in XRD patterns of our film samples by persuading that particles are in nanometer range. Particles sizes less than 5nm can't be detected using XRD.

| Concentration | Band gap energy(eV) |
|---------------|---------------------|
| 0%            | 3.27                |
| 1%            | 3.26                |
| 2%            | 3.25                |
| 3%            | 3.24                |
| 4%            | 3.24                |
| 5%            | 3.18                |

Table 2: Calculated values of energy gap at different Cu concentrations.

**4. Conclusion:**

Films were prepared using a low cost spin coating technique. Below the absorption edge, the transmittance of samples increases with Cu concentration of ZnO thin film sample. The absorption edge can be observed around 380 nm. Absorbance, reflectance and refractive index decrease with Cu concentration of ZnO sample below this wavelength. However, a systematic variation of transmittance, absorbance, reflectance or refractive index with wavelength couldn't be observed above the absorption edge. The thickness of the samples calculated from refractive index versus wavelength graphs varies from 108.3 to 138 nm depending on the amount of the solution spread on the substrate. Energy gaps of the samples determined from the graph of $(\alpha h\nu)^2$ versus photon energy $(h\nu)$ gradually decreases with Cu concentration doped with ZnO. The reason for the variation of the energy gap could be the electron concentration dependence of the band gap shift.

**References:**

1. M. Caglar, Y. Caglar and S. Ilican, 2006. The determination of the thickness and optical constants of the ZnO crystalline thin film by using envelope method. Journal of Optoelectronics and Advanced Materials 8(4), 1410-1413.

2. H.F. Hussein, Ghufran Mohammad Shabeeb and S. Sh. Hashim, 2011. Preparation ZnO thin film by using sol-gel processed and determination of thickness and study optical properties. Journal of Materials and Environmental Science 2(4), 423-426.

3. S. Ilican, Y. Caglar and M. Caglar, 2008. Preparation and characterization of ZnO thin films deposited by sol-gel spin coating method. Journal of Optoelectronics and Advanced Materials 10(10), 2578-2583.

4. A. Mosbah, A. Moustaghfir, S. Abed, N. Bouhssira, M.S. Aida, E. Tomasella and M. Jacquet, 2005. Comparison of the structural and optical properties of zinc oxide thin films deposited by dc and rf sputtering and spray pyrolysis. Surface and Coatings Technology 200, 293-296.

5.  Wang Zhao-yang, Hu Li-zhong, Zhao Jie, Sun Jie and Wang Zhi-jun, 2005. Effect of the variation of temperature on the structural and optical properties of ZnO thin films prepared on Si(111) substrates using PLD.  Vacuum 78(1), 53-57.

6.  A. Mosbah, S. Abed, N. Bouhssira, M.S. Aida and E. Tomasella, 2006. Preparation of highly textured surface ZnO thin films. Materials Science and Engineering B. 129, 144-149.

7. Said Benramache, Boubaker Benhaoua and Foued Chabane, 2012. Effect of substrate temperature on the stability of transparent conducting cobalt doped ZnO thin films. Journal of Semiconductors 33(9), 093001–1.

8.  Said Benramache, Boubaker Benhaoua and Hamza Bentrah, 2013.Preparation of transparent conductive ZnO:Co and ZnO: In thin films by ultrasonic spray method. Journal of Nanostructure in chemistry 3, 54.

9. T. Yamada, T. Nebiki, S. Kishimoto, H. Makino, K. Awai, T. Narusawa and T. Yamamoto, 2007. Dependence of structural and electrical properties on thickness of polycrystalline Ga-doped ZnO thin films prepared by reactive plasma deposition. Superlattices and Microstructures 42, 68-73.

10. J.R. Duclère, M. Novotny, A. Meaney, R. O'Haire, E. McGlynn, M.O. Henry and J.P. Mosnier, 2005. Properties of Li -, P- and N-doped ZnO thin films prepared by pulsed laser deposition. Superlattices and Microstructures 38, 397-405.

11. A. Djelloul, M.S. Aida and J. Bougdira, 2010. Photoluminescence, FTIR and X-ray diffraction studies on undoped and Al-doped ZnO thin films grown on polycrystalline α–alumina substrates by ultrasonic spray pyrolysis. Journal of Luminescence 130, 2113-2117.

12. H. Kavak, E.S. Tuzemen, L.N. Ozbayraktar and R. Esen, 2009. Optical and photoconductivity properties of ZnO thin films grown by pulsed filtered cathodic vacuum arc deposition. Vacuum 83(3), 540-543.

13. Yinzhen Wang and Benli Chu, 2008. Structural and optical properties of ZnO thin films on (111) CaF$_2$ substrates grown by magnetron sputtering. Superlattices and Microstructures 44, 54-61.

14. Zhiyun Zhang, Chonggao Bao, Wenjing Yao, Shengqiang Ma, Lili Zhang and Shuzeng Hou, 2011. Influence of deposition temperature on the crystallinity of Al-doped ZnO thin films at glass substrates prepared by RF magnetron sputtering. Superlattices and Microstructures 49, 644-653.

15. B.L. Zhu, X.Z. Zhao, F.H. Su, G.H. Li, X.G. Wu, J. Wu and R. Wu, 2010. Low temperature annealing effects on the structure and optical properties of ZnO films grown by pulsed laser deposition. Vacuum 84(11), 1280-1286.

16. P. Samarasekara, A.G.K. Nisantha and A.S. Disanayake, 2002. High Photo-Voltage Zinc Oxide Thin Films Deposited by DC Sputtering. Chinese Journal of Physics 40(2), 196-199.

17. P. Samarasekara, 2010. Characterization of Low Cost p-$Cu_2$O/n-CuO Junction. Georgian Electronic Scientific Journals: Physics 2(4), 3-8.

18. P. Samarasekara, 2009. Hydrogen and Methane Gas Sensors Synthesis of Multi-Walled Carbon Nanotubes. Chinese Journal of Physics 47(3), 361-369.

19. P. Samarasekara and N.U.S. Yapa, 2007. Effect of sputtering conditions on the gas sensitivity of Copper Oxide thin films. Sri Lankan Journal of Physics 8, 21-27.

20. K. Tennakone, S.W.M.S. Wickramanayake, P. Samarasekara and, C.A.N. Fernando, 1987. Doping of Semiconductor Particles with Salts. Physica Status Solidi (a)104, K57-K60.

21. P. Samarasekara and Udara Saparamadu, 2013. Easy axis orientation of Barium hexa-ferrite films as explained by spin reorientation. Georgian Electronic Scientific Journals: Physics 1(9), 10-15.

22. P. Samarasekara and Udara Saparamadu, 2012. Investigation of Spin Reorientation in Nickel Ferrite Films. Georgian electronic scientific journals: Physics 1(7): 15-20.

23. P. Samarasekara and N.H.P.M. Gunawardhane, 2011. Explanation of easy axis orientation of ferromagnetic films using Heisenberg Hamiltonian. Georgian electronic scientific journals: Physics 2(6): 62-69.

24. P. Samarasekara. 2008. Influence of third order perturbation on Heisenberg Hamiltonian of thick ferromagnetic films. Electronic Journal of Theoretical

Physics 5(17): 227-236.

25. P. Samarasekara and Udara Saparamadu, 2013. In plane oriented Strontium ferrite thin films described by spin reorientation. Research & Reviews: Journal of Physics-STM journals 2(2), 12-16.

26. P. Samarasekara, 2008. Four layered ferromagnetic ultra-thin films explained by second order perturbed Heisenberg Hamiltonian. Ceylon Journal of Science 14, 11-19